\documentclass[aps,prd,twocolumn,showpacs,groupedaddress,floatfix,nofootinbib]{revtex4}
\usepackage{graphicx}  
\usepackage{dcolumn}   
\usepackage{bm}        
\usepackage[english]{babel}
\usepackage{amsfonts,amsmath,amssymb,mathrsfs}
\usepackage{epstopdf}
\usepackage{subfigure}
\usepackage{color}
\usepackage{hyperref}

\def\a{\alpha}
\def\r{\rho}
\def\s{\sigma}
\def\t{\tau}
\def\m{\mu}
\def\n{\nu}
\def\k{\kappa}
\def\th{\theta}
\def\g{\gamma}\def\G{\Gamma}
\def\L{\Lambda}\def\l{\lambda}
\def\D{\Delta}
\def\la{\langle}
\def\ra{\rangle}
\def\o{\omega}\def\O{\Omega}
\def\d{\delta}
\def\p{\partial}

\def\half{\textstyle{\frac{1}{2}}}

\def\bdoc{\begin{document}}
\def\edoc{\end{document}}

\def\beq{\begin{equation}}
\def\eeq{\end{equation}}
\def\bea{\begin{eqnarray}}
\def\eea{\end{eqnarray}}
\def\ben{\begin{enumerate}}
\def\een{\end{enumerate}}
\def\la{\langle}
\def\ra{\rangle}
\def\a{\alpha}
\def\b{\beta}
\def\g{\gamma}\def\G{\Gamma}
\def\d{\delta}\def\D{\Delta}
\def\e{\epsilon}
\def\z{\zeta}

\def\th{\theta}
\def\k{\kappa}
\def\l{\lambda}
\def\m{\mu}
\def\n{\nu}
\def\o{\omega}
\def\p{\pi}
\def\r{\rho}
\def\s{\sigma}
\def\t{\tau}
\def\S{\Sigma }
\def\gsim{\; \raisebox{-.8ex}{$\stackrel{\textstyle >}{\sim}$}\;}
\def\lsim{\; \raisebox{-.8ex}{$\stackrel{\textstyle <}{\sim}$}\;}
\def\lessim{\lsim}
\def\loc{{\rm local}}
\def\vm{v_{\rm max}}
\def\bh{\bar{h}}
\def\del{\partial}
\def\nab{\nabla}
\def\half{{\textstyle{\frac{1}{2}}}}
\def\fourth{{\textstyle{\frac{1}{4}}}}

\def\bD{{\bf D}}
\def\bE{{\bf E}}
\def\bF{{\bf F}}
\def\bB{{\bf B}}
\def\bP{{\bf P}}
\def\bV{{\bf v}}
\def\bv{{\bf v}}
\def\bx{{\bf x}}
\def\by{{\bf y}}
\def\bz{{\bf z}}
\def\ba{{\bf a}}
\def\bd{{\bf d}}
\def\bs{{\bf s}}
\def\bn{{\bf n}}
\def\bp{{\bf p}}

\def\O{\Omega}

\def\br{{\bf r}}
\def\bnab{{\bf \nab}}

\def\tE{\tilde{E}}
\def\tL{\tilde{L}}
\def\Horava{Ho\v{r}ava }

\begin{document}

\title{Cosmic alignment of the aether}
\author{Isaac Carruthers and Ted Jacobson}
\affiliation{Center for Fundamental Physics,  University of Maryland, College Park, Maryland 20742-4111, USA}
\date{\today} 
\begin{abstract}
In Einstein-aether theory and Horava gravity,
a timelike unit vector is coupled to the spacetime metric.
It has previously been shown that in an exponentially expanding
homogeneous, isotropic background, small perturbations of the vector 
relax back to the isotropic frame. Here we investigate large
deviations from isotropy, maintaining homogeneity. We find that,
for generic values of the coupling constants, 
the aether and metric relax to the isotropic configuration
if the initial aether hyperbolic boost angle and its time derivative
in units of the cosmological constant are less than something of
order unity. For larger angles or angle derivatives, 
the behavior is strongly dependent
on the values of the 
coupling constants. Generally there is runaway behavior, in which the 
anisotropy increases with time, and/or singularities occur.

\end{abstract}  

\pacs{04.50.Kd, 11.30.Cp, 98.80.Cq}

\maketitle

\section{Introduction}
 
When the phenomenology of theories with a 
preferred frame is studied, 
it is generally assumed that this frame
coincides, at least roughly, with the cosmological
rest frame defined by the Hubble expansion of the 
universe. Observations place
strong bounds on frame dependent effects which would
presumably grow with the relative velocity of the 
preferred frame and the velocity of the earth (which 
moves at the ``low" speed of $10^{-3} c$ relative to the 
Hubble frame).
In a particular theory with preferred frame effects, 
the dynamical alignment of the frame (or frames) can be studied
to determine stability of cosmic alignment, as well as
to characterize the range of initial conditions that 
could be expected to naturally align. 

In this paper we examine this question in the 
case of Einstein-aether theory\cite{Jacobson:2008aj} and in the IR
limit of (extended) Horava gravity\cite{Horava:2009uw,Blas:2009qj}.
Einstein-aether theory just consists of general relativity coupled,
at second derivative order, to 
a dynamical timelike unit vector field $u^a$, the aether.
In Horava gravity, the aether vector is assumed to be 
hypersurface-orthogonal, i.e. it is the unit normal to 
level sets of a scalar time function.
Various forms of ``Horava gravity" have been discussed in
the literature. Here we refer exclusively to the one related
to Einstein-aether theory as just explained. (This corresponds 
to the so-called  ``non-projectable" version, where the lapse 
function $N$ is an arbitrary function of spacetime, 
and includes in the Lagrangian a term proportional to 
the square of the gradient of $\ln N$.) 
Every hypersurface-orthogonal
Einstein-aether solution is a Horava solution. 
All the solutions to be considered in this paper are of this type.

The alignment of the aether has been studied before
in the context of linearized perturbations. 
The question was first addressed, indirectly, by Lim\cite{Lim:2004js},
who found that all perturbations
of the aether decay exponentially
in a de Sitter background. In particular, this result
applies to the homogeneous modes.
Subsequent work\cite{Li:2007vz,ArmendarizPicon:2010rs}
confirmed this result, but in Ref.~\cite{ArmendarizPicon:2010rs}
it was found that under some circumstances,
after inflation, velocity perturbations
might grow to be ``mildly relativistic" and could still
possibly be compatible with observations.
In all these analyses, it is assumed that the 
aether is aligned in a background solution,
and the behavior of perturbations is studied.

Kanno and Soda (KS) approached the question 
from a different point of view. 
In the Appendix of Ref.~\cite{Kanno:2006ty} they examined 
homogeneous but anisotropic 
solutions in the presence of a positive cosmological
constant, with three orthogonal principal 
directions of expansion, and
with the aether tilted in one of the
principal directions. [This corresponds to 
Bianchi type I (Kasner-like) symmetry.]
They showed that, to linear order in the anisotropy, the system
relaxes exponentially to the isotropic, de Sitter solution.
Since their analysis was carried out just to linear order
in the anisotropy, 
it is in fact just a special case of the above-mentioned 
perturbative treatments. 

In this paper we adopt precisely 
the setting of the KS analysis
but we include the full nonlinear dynamics. 
We characterize the range of initial data that
relax to an isotropic solution. 
Generically, the aether aligns provided the
initial boost angle and its time derivative
in units of the cosmological constant are 
less than something
of order unity. The precise stability
bounds depend on the values of the 
coupling parameters in the Lagrangian 
defining the theory.

\section{Bianchi type I Einstein-aether cosmology}

Einstein-aether theory 
is general relativity (GR) 
coupled to a dynamical timelike unit vector field.
In terms of the metric $g_{ab}$ of signature $(+{-}{-}{-})$ 
and the unit vector field $u^a$ it is defined by the action
\begin{equation} \label{Lae}
S = \frac{-1}{16\pi G}\int d^4x\, \sqrt{-g} \left( R+2\Lambda + K^{ab}_{\phantom{ab}mn} \nabla_a u^m \nabla_b u^n  \right)
\end{equation}
where $R$ is the Ricci scalar, $\L$ is a cosmological constant, 
and the tensor $K^{ab}_{\phantom{ab}mn}$ is given by
\begin{equation}
K^{ab}_{\phantom{ab}mn} = c_1 g^{ab} g_{mn} + c_2 \delta^a_m \delta^b_n + c_3 \delta^a_n \delta^b_m + c_4 u^a u^b g_{mn},
\end{equation}
and $c_1, \ldots, c_4$ are dimensionless coupling parameters that define the theory.
Since $u^m$ is constrained to be a unit vector, the action need 
only be stationary under variations orthogonal to the aether, 
$u_m \d u^m=0$.  
In this paper for simplicity we omit any matter couplings,
since the cosmological constant suffices to 
source the overall expanding solution and 
it models the conditions that would have pertained in 
an inflationary early universe. It would be straightforward to
add radiation or matter or some form of quintessence to the 
model. 

\subsection{Bianchi type I symmetry}
Following KS, we specialize to Bianchi type I spacetimes,
i.e. to metrics that are homogeneous and spatially flat,
with three commuting translation symmetries,
\bea\label{BianchiI}
ds^2 &=& N^2(t) dt^2 - e^{2\a(t)}\bigl[e^{-4\s_+(t)} dx^2 \nonumber\\
&&+ e^{2\s_+(t)}(e^{2\sqrt{3}\s_-(t)}dy^2
+e^{-2\sqrt{3}\s_-(t)} dz^2)\bigr].
\eea
We also assume that the aether vector is 
tilted only in the $x$-direction, 
\beq\label{u}
u= \frac{1}{N(t)}\cosh\theta(t)\, \del_t 
+ e^{-\a(t)+2\s_+(t)}\sinh\theta(t)\, \del_x.
\eeq
The hyperbolic angle $\th$ measures the boost of the 
aether relative to the rest frame of the homogeneous,
flat spatial sections, i.e. the ``homogeneous frame".
The metric is determined by four functions. The lapse $N(t)$ 
specifies the flow rate of proper time with respect to $t$
in the homogeneous frame.
Co-moving lengths $L$ 
in the $x$, $y$, and $z$ directions all have different expansion
rates, $\dot L/L$. The sum of these is $3\dot\a$,
which is also the fractional rate of change of 
co-moving volume, $\dot V/V$. 
The quantity $2\sqrt{3}\dot\s_-$ is the 
difference between the expansion rates in the 
two transverse directions $y$ and $z$, 
while $3\dot\s_+$ is the difference between the 
average of these and the rate in the $x$ direction.

The vector field (\ref{u}) is in effect just two dimensional; hence, 
like all such vector fields, it is hypersurface orthogonal.
According to the analysis of Ref.~\cite{Jacobson:2010mx}, this means that 
the solutions to the field equations discussed here are also
solutions to the field equations of Horava gravity.
Hence our results apply to the cosmology of that theory as well.
The hypersurface orthogonality also means\cite{Eling:2006df} 
that the
action is unchanged under
$c_1\rightarrow c_1+\d$, $c_3\rightarrow c_3-\d$, and $c_4\rightarrow c_4-\d$,
so the system depends on these three coupling
parameters only through the two invariant combinations
$c_1+c_4$ and $c_1+c_3$. For notational compactness
we shall make use of these quantities, and
also drop the subscript 2 on $c_2$:  
\beq
a=c_1+c_4,  \quad b=c_1+c_3, \quad c=c_2. 
\eeq
[These parameters correspond respectively to the
parameters $\a,\b,\l'$ of the action for
Horava gravity, Eq. (5.72) in Ref.~\cite{Blas:2010hb}.]

When the fields have this symmetry structure, 
the action takes the form  (up to a total derivative)
\beq\label{action}
S= \frac{1}{16\pi G}\int dt \, e^{3\a}\Bigl(\frac{1}{2N} H_{ij}(\theta)\dot{q}^i\dot{q}^j
-2N\Lambda\Bigr),
\eeq
where $q^i\leftrightarrow (\theta, \a, \s_+,\s_-)$.
Here and below the time dependence of the dynamical
variables is implicit, the dot denotes derivative with respect to $t$, 
and indices $i,j,\dots$  label the four 
dynamical variables. 
The nonzero components of the symmetric array $H_{ij}$ are given by 
\bea\label{H}
H_{\th\th}&=&2\bigl(b+c + (a-b-c)\cosh^2\th\bigr)\nonumber\\
H_{\th\a}&=& 2(a-b-3c)\cosh\th\, \sinh\th\nonumber\\
H_{\th+}&=&4(-a+b)\cosh\th\, \sinh\th\nonumber\\
H_{\a\a}&=&2\bigl(-6 -a+ (a-3b-9c)\cosh^2\th\bigr)\nonumber\\
H_{\a+}&=&-4a\sinh^2\th\nonumber\\
H_{++}&=&4\bigl(3-2a+(2a-3b)\cosh^2\th\bigr) \nonumber\\
H_{--}&=&12(1-b\cosh^2\th)
\eea
The $\th$ dependence of $H_{ij}(\th)$ will also be suppressed.
The dynamics is symmetric under the inversion
$(\th,\dot\th)\rightarrow(-\th,-\dot\th)$.

A key to the general behavior of solutions is the invertibility 
of $H_{ij}$, whose determinant can be written in the form
\bea\label{detH}
\det H &=&-1728a(1-b)^2(2+b+3c)\nonumber\\
&&\hspace{-15mm}\times\bigl(v_0^2 +(1-v_0^2)\cosh^2\theta\bigr) 
\bigl(v_2^2 +(1-v_2^2)\cosh^2\theta\bigr) 
\eea
with
\beq\label{speeds}
v_2^2 = \frac{1}{1-b},\qquad v_0^2 = \frac{(b+c)(2-a)}{a(1-b)(2+b+3c)}.
\eeq
The constants
$v_0$ and $v_2$ are, in fact, the speeds of the 
spin-0 and spin-2 modes linearized around flat space
(defined with respect to the background aether frame), 
in both Einstein-aether theory\cite{Jacobson:2008aj}
and Horava gravity. Note that the $v_0$ factor is
proportional to $H_{--}$. 

It may initially be  surprising that the 
spin-0 and spin-2 perturbations 
play any role in a homogeneous
cosmology.  But while the metric
(\ref{BianchiI}) is homogeneous on constant
$t$ surfaces, it has $x$ dependence on surfaces
orthogonal to the aether. 
In fact, the form of (\ref{detH}) can be understood 
from simple kinematic 
considerations as follows.

The 
linearized, diagonalized,  
action for a mode $\psi$ with speed $v$
is proportional to $g^{ab}_{(v)}\partial_a\psi\partial_b\psi$,
where $g^{ab}_{(v)}\propto u^au^b + v^2(g^{ab} - u^au^b)$
is the effective metric for that mode. Since we consider
only homogeneous fields, which depend on $t$ alone,
the only component that enters is 
$g^{tt}_{(v)}\propto \bigl(v^2 g^{tt} +(1-v^2)u^tu^t\bigr)
\propto \bigl(v^2 +(1-v^2)\cosh^2\th\bigr)$. The symmetry
and hypersurface orthogonality of the aether 
permit only one spin-2 mode
and the spin-0 mode. The spin-2 mode is governed
by the shear $\sigma_-$, which describes the gravitational 
wave mode transverse to the tilt and with polarization aligned with 
the $y$ and $z$ symmetry axes. This explains the form of the
determinant (\ref{detH}).

If the coupling constants $(a,b,c)$ are such that
one of the mode speeds exceeds unity, then there
is a value of $\th$ for which the determinant of 
$H_{ij}$ vanishes, corresponding to the
condition $g^{tt}_{(v)}=0$. When the aether reaches this 
hyperbolic tilt angle, the propagation cone of that
mode becomes tangent to the 
constant $t$ surface, so that
becomes a valid constant phase surface 
for the mode. In other words, the mode propagates
instantaneously on the constant $t$ surface.
Beyond this aether tilt the kinetic energy of 
the mode becomes negative, so the 
system is unstable.
In section \ref{generic} we shall discuss the
implications of this phenomenon 
for the cosmological dynamics.

%

\subsection{Equations of motion for $\Lambda=0$}

Although our main interest is in the case of exponential expansion
driven by a positive cosmological constant, we begin by looking first
at the simpler case of vanishing $\Lambda$. 
Then variation with respect to the lapse $N$ yields the initial
value constraint
\beq\label{constraintq0}
H_{ij}\dot{q}^i\dot{q}^j=0.
\eeq
As usual in general relativity, 
if this constraint is satisfied at one time, then the
rest of the equations of motion imply that it
remains satisfied for all time.
We can choose the nontrivial lapse
\beq \label{N0}
N=e^{3\a}
\eeq
to eliminate the $\a$ dependence in the action. 
Then the dynamics becomes that of 
affinely parametrized null geodesics on the configuration
space $(\th,\a,\s_+,\s_-)$ with respect to the metric 
$H_{ij}$ that depends only on $\th$. 

It is convenient to define momenta by 
\beq\label{p}
p_i = H_{ij} \dot{q}^j,
\eeq
which can be solved for 
the velocities, 
\beq\label{q}
\dot{q}^i= H^{ij} p_j,
\eeq
when the inverse $H^{ij}$ of $H_{ij}$ exists.
In terms of the momenta, the constraint (\ref{constraintq0}) reads
\beq\label{zeroLconstraint}
H^{ij}p_ip_j=0.
\eeq
Since $H_{ij}$ depends only on $\th$,
the momenta $p_\a$ and $p_\pm$ are conserved.
Moreover, the constraint  
(\ref{zeroLconstraint}) is 
a quadratic equation in $p_\th$ that can be 
solved for $p_\th(\th; p_\a,p_\pm)$ (there are generically two roots or none).
Having solved three of the four evolution equations, as well as the constraint
equation, the fourth evolution equation, for $p_\th$, is now redundant.
The dynamics for $\th$ is thus reduced to a first order differential equation,
\beq\label{dth}
\dot{\th}=H^{\th k}p_k =: F(\th; p_\a,p_\pm).
\eeq
Once the evolution of $\th$ is known, the remaining
variables $\a$ and $\s_\pm$ 
are determined by integration of the first
order equation (\ref{q}). The character of the evolution
of $\th$ can be seen by inspection of a plot of the
graph of the function $F$ defined in (\ref{dth}).

\subsection{Equations of motion for $\Lambda\ne0$}
\label{eomLne0}

For nonvanishing $\Lambda$,
the variation with respect to the lapse $N$ yields the initial
value constraint,
\beq\label{constraintq}
H_{ij}\dot{q}^i\dot{q}^j+4\Lambda=0.
\eeq
Because of the $\L$ term in the action,
the lapse (\ref{N0}) is no longer 
the most convenient, and it is simpler to 
just use
\beq \label{N1}
N=1.
\eeq
The Euler-Lagrange equations with this gauge choice are
\beq\label{ELeq}
\frac{d}{dt}\left(e^{3\a}H_{ij}\dot{q}^j\right)-\del_i\Bigl(e^{3\a}\bigl(\half H_{kl}\dot{q}^k\dot{q}^l
-2\Lambda\bigr)\Bigr)=0.
\eeq
The individual components $i=\th,\a,\pm$ read
\bea
\frac{d}{dt}\left(e^{3\a}H_{\th j}\dot{q}^j\right)&=
&\half e^{3\a}H_{ij,\th}\dot{q}^i\dot{q}^j \\
\frac{d}{dt}\left(e^{3\a}H_{\a j}\dot{q}^j\right)&=&-12\Lambda e^{3\a}\label{alphaeqn}\\
\frac{d}{dt}\left(e^{3\a}H_{\pm j}\dot{q}^j\right)&=& 0,
\eea
where in (\ref{alphaeqn}) the constraint (\ref{constraintq}) was used.

Again, it is sometimes convenient to express the field equations
in terms of the ``momenta" (\ref{p}). 
(These are not precisely the conjugate momenta anymore 
since the factor $e^{3\a}$ is not included, but we will nevertheless
refer to them as momenta.) 
Then the constraint (\ref{constraintq}) takes the form
\beq\label{constraint}
H^{ij}p_ip_j+4\Lambda=0,
\eeq
and the equations of motion become
\bea
\dot{p}_\th&=&-3\dot\a\, p_\th 
-\half H^{ij}_{\phantom{ij},\th}\, p_ip_j\label{eomth}\\
\dot{p}_\a&=&-3\dot\a\, p_\a -12\Lambda \label{eoma}\\
\dot{p}_\pm&=& -3\dot\a\, p_\pm\label{eompm}
\eea
Using (\ref{q}) we can express $\dot\a$ in terms of 
$\th$ and the momenta, 
\beq\label{alphadot}
\dot\a=H^{\a k} p_k.
\eeq
Moreover, we can solve the quadratic constraint 
equation (\ref{constraint}) for 
$p_\a$, so that 
$\a$ and $p_\a$ 
can be eliminated completely from the 
dynamical system.

\section{Limiting cases}
In this section we discuss various special cases and limits
of the theory. 

\subsection{General relativity}

If we reduce to the pure GR case,
$\th$ is not present, and $a=b=c=0$. 
Then $H_{ij}$ is diagonal and constant, 
with $H_{++}=H_{--}=-H_{\a\a}=12$. 
The constraint (\ref{constraint}) then becomes
\beq
-p_\a^2+p_+^2+p_-^2 =-48\Lambda.
\eeq
The only isotropic ($p_\pm=0$) solutions
are Minkowski spacetime (with $\L=0$) and 
de Sitter spacetime (with $\dot\a=\sqrt{\L/3}$, $\L>0$).
In the anisotropic case, $\L=0$ yields the Kasner solutions,
and $\L\ne0$ yields a generalization of  
those.

\subsection{$\theta=0$ solutions}

Next we characterize the solutions in which 
$\th=0$ for all times, i.e. in which
the aether remains orthogonal
to the constant $t$ homogeneity surfaces. 
Then, although the aether has no 
dynamics,  its 
couplings in the action (\ref{Lae})  
still contribute to the field equations
and we therefore have something different from GR.

There are no
terms linear in $\th$ or $\dot{\th}$ alone
in the action  (\ref{action}), and terms of quadratic
or higher order in these quantities 
will obviously not contribute to the equations of
motion if $\th=\dot\th=0$. If this 
condition holds initially it is therefore preserved for all
time, and for characterizing 
these solutions it is consistent to 
simply set $\th=0$ in the action. Then
$H_{ij}$ is diagonal and constant, the relevant components being
\bea\label{Hzerotheta}
H_{\a\a}&=&-6(2 +b+3c)\nonumber\\
H_{++}&=&12(1-b) \nonumber\\
H_{--}&=&12(1-b).
\eea
In the isotropic case $\s_\pm=0$, the system is then
equivalent to GR with a rescaled cosmological value
of Newton's  
constant~\cite{Mattingly2001,Carroll:2004ai}, 
$G_{\rm cosmo}=G/(1+(b+3c)/2)$,
and with $\Lambda$ replaced by $\Lambda'=\Lambda/(1+(b+3c)/2)$.
This isotropic solution
is the spatially flat slicing of 
de Sitter spacetime with Hubble constant $H=\sqrt{\Lambda'/3}$.
The aether becomes singular because of infinite stretching 
on the past horizon.
In the anisotropic case, interestingly,
there is no equally simple relation to GR:
{\it the presence of the isotropic aether induces 
different rescalings of the
kinetic energy associated with expansion and shear.}

\subsection{Linearized anisotropy}

If one drops all terms in the action (\ref{action}) of
higher than quadratic order in the
anisotropic coordinates $\th$ and $\s_\pm$, the 
$H_{ij}$ array reduces to the nonzero elements 
\bea\label{Hlin}
H_{\th\th}&=&2a\nonumber\\
H_{\th\a}&=& 2(a-b-3c)\th\nonumber\\
H_{\a\a}&=&-6(2 +b+3c)+2(a-3b-9c)\th^2\nonumber\\
H_{++}&=&12(1-b) \nonumber\\
H_{--}&=&12(1-b).
\eea
Keeping only linear order terms in the 
anisotropy, the equation of motion for $\th$ 
reduces to 
\beq\label{linearized}
\ddot\th + 3\dot\a \, \dot\th + 2\dot{\a}^2\, \th = 0.
\eeq
To zeroth order in the anisotropy the solution to the 
constraint (\ref{constraintq}) is, 
\beq
\dot\a^2 = \frac{\L}{3[1+(b+3c)/2]}
\eeq
and $\dot\a$ in (\ref{linearized}) can be replaced by this value.
Then (\ref{linearized})  
is the equation found by KS~\cite{Kanno:2006ty}. They pointed out that 
the coefficient of $\th$ is positive provided the effective 
gravitational coupling is positive, in which case this is
the equation of a damped harmonic oscillator.
In fact, the oscillator is overdamped,
with eigenmode decay rates 
$\dot\a$ and $2\dot\a$. This
implies that $\theta$ relaxes to zero as the universe expands.


%

\subsection{$p_\pm=0$ solutions}
\label{zeropm}


The equation of motion
(\ref{eompm}) is solved for $p_\pm$ by 
\beq\label{pdecay}
p_\pm(t)=e^{-3\a(t)}\, p_{\pm,0},
\eeq
where $p_{\pm,0}$ is an integration
constant. Using this, the remaining
equations of motion (\ref{eomth},\ref{eoma}) 
involve only the variables $(\th, p_\th, \a, \dot\a, p_\a)$.
Moreover, as mentioned at the end of section \ref{eomLne0},
Eq.\ (\ref{alphadot}) can be used to eliminate
$\dot\a$, and the constraint can be solved for 
$p_\a$. However this requires that
the function $\a(t)$ be determined by the 
previous values of $(\th, p_\th, \a)$ via $\int^t dt'\, \dot\a(t')$,
which does not yield an evolution equation that is local in time.

If the co-moving volume is 
expanding then, according to 
(\ref{pdecay}), $p_\pm(t)$ is driven
to zero. It is therefore a useful limiting case
to set $p_\pm=0$ from the beginning. Then
there is no remaining $\a(t)$ dependence, and the
system can be reduced to 
the $\th$ degree of freedom alone. We now
explain in detail how this is achieved.

We assume now that $p_\pm=0$.
Since $H^{-k}=0$ except for $k=-$, it follows from (\ref{q})
that $\dot{\s}^-=0$. Thus, the two transverse dimensions
must have the
same expansion rates. In contrast, it
does {\it not} follow that $\dot{\s}^+=0$,
since we have in this case
\beq\label{sdot}
\dot{\s}^+ =H^{+ \th}p_\th + H^{+ \a}p_\a.
\eeq
As might be expected, it is 
inconsistent  
for the metric to be isotropic ($\s^\pm=0$)
when the aether is tilted ($\th\ne0$).
(However, note that
$\dot{\s}^+$ is of second order in the $\theta$ anisotropy.)
The expansion rate in the tilt direction must generally
differ from that in the transverse direction. 
Eliminating $\dot{\s}^+$ via (\ref{sdot}),
the system reduces to the variables ($\th$, $p_\th$, $\a$, $p_\a$).

One can write this system
in terms of the velocities, i.e.\ in terms of the
variables
$(\th,  \dot{\th}, \a, \dot{\a})$, by
using the constraint (\ref{constraintq}) to solve 
for $\dot{\s}^+$
in terms of $(\th, \dot{\th}, \dot{\a})$, and substituting
that into the Euler-Lagrange equation (\ref{ELeq}).
But for the purpose of making a $(\th, \dot\th)$ 
phase portrait
of the evolution, it appears more neat
to organize the equations as follows.

The idea is to 
solve for the momenta $(p_\th, p_\a)$
in terms of the velocities $(\dot{\th}, \dot{\a})$, and then
to use the equations that were expressed in terms of momenta.
To this end, we introduce capital indices $A,B,...$ to
refer to the two coordinates $\th$ and $\a$, 
we define the contravariant tensor $h^{AB}$
to be the restriction of $H^{ij}$, 
\beq
h^{AB}\equiv H^{AB},
\eeq
and we denote by $h_{AB}$ the inverse of $h^{AB}$.
Then from (\ref{q}) we have
\beq\label{qA}
\dot{q}^A= h^{AB}p_B,
\eeq
which can be inverted to yield
\beq\label{pA}
p_A=h_{AB}\dot{q}^B.
\eeq
Using (\ref{pA})
the constraint (\ref{constraint}) becomes
\beq
h_{AB}\dot{q}^A\dot{q}^B+4\Lambda=0,
 \eeq
which can be solved as a quadratic equation for $\dot\a(\th,\dot{\th})$,
thus eliminating $\dot\a$. Explicitly, we have
\beq\label{adot}
\dot{\a}=\frac{-h_{\th\a}\dot{\th}
\pm\sqrt{(h_{\th\a}\dot{\th})^2-h_{\a\a}(h_{\th\th}\dot{\th}^2+4\Lambda})}{h_{\a\a}}.
\eeq
There are generically two solutions or no solutions.
If $h_{\a\a}(h_{\th\th}\dot{\th}^2+4\Lambda)<0$
there are two solutions, 
one in which the volume is expanding ($\dot\a>0$) and the other 
in which it is contracting ($\dot\a<0$). Note that for 
$\dot\th=0$ the solutions are simply $\dot\a=\pm\sqrt{-4\Lambda/h_{\a\a}}$.
As $h_{\a\a}\rightarrow 0^-$ this diverges, and no solution exists
when $h_{\a\a}>0$.

The remaining task is to find an equation for $\ddot \th$.
For this purpose we can use (\ref{q}) to write 
$\dot{\th}=h^{\th B}p_B$; hence, 
\beq
\ddot{\th}=h^{\th B}{}_{,\th}\, \dot{\th}\, p_B+h^{\th B}\dot{p}_B.
\eeq
Then using (\ref{pA}) and the equations of motion
(\ref{eomth},\ref{eoma}) we find, after some manipulation,
\bea
\ddot{\th}&=&-3\dot{\a}\dot{\th}-12\Lambda h^{\th\a}-h^{\th A}h_{AB,\th}\, \dot{q}^B\dot{\th}\\
&&+\half h^{\th\th}h_{AB,\th}\, \dot{q}^A\dot{q}^B.
\eea
Together with (\ref{adot}) this yields a dynamics reduced to just the 
$\th$ degree of freedom, which can be visualized in a phase portrait.

\section{Generic behavior}\label{generic}
We have seen that, in a small enough neighborhood
of $\th=0$, the dynamics relaxes to the $\th=0$ case,
provided the values of $a$, $b$, and $c$ are such that the 
effective gravitational coupling constant is positive.
On the other hand, once $\th$ is sufficiently large,
the character of the dynamical system can obviously
change dramatically because of the growth of the
hyperbolic trigonometric functions in the components
of $H_{ij}$ (\ref{H}). 

A general feature mentioned earlier is that $\det H$ (\ref{detH})
vanishes if either the spin-0 or spin-2 
propagation cone is tangent to the constant
$t$ surface. The conditions determining these angles
can be expressed as
\beq\label{degeneracy}
\coth\th_0=v_0, \quad\coth\th_2=v_2
\eeq
where $v_{0,2}$ are the mode speeds (\ref{speeds}).
At either of these angles $H_{ij}$ is 
not invertible, so the equation of motion (\ref{ELeq}) cannot 
be solved for $\ddot{q}^i$. As such a value of $\theta$ is 
approached, at least one second derivative component would
generally diverge. Hence generically 
there can be no smooth 
evolution across the degenerate values of $\th$. 
The dynamics may run into a singularity there, 
or it may ``bounce" before reaching such a value of $\th$.

There is a solution $\th_*$ to each of the
equations in (\ref{degeneracy}) 
as long as the corresponding mode speed $v$ is 
greater than unity. The {\it larger} of the mode
speeds defines the {\it smaller} of the critical angles. 
The critical angles
are of order unity unless $v$ is either very large or
very close to 1. In these limits
we have
\beq
\th_*\approx \left\{\begin{array}{cl}
1/v&\mbox{for}\quad v\gg1\\
-\half\ln(v-1)\quad &\mbox{for}\quad v-1\ll1
\end{array}\right.
\eeq
In particular, the degenerate value $\th_2$ is real only
if $0<b\le1$, and
is of order unity unless $b$ is 
very close to either 0 or 1. For instance,
for $b=0.01$, 0.9, or 0.99, we have
$\th_2 \simeq 3$, 0.3, or 0.1 respectively. 
The degenerate value $\th_0$ is 
of order unity for 
generic values of $a,b,c$ with no
large hierarchy amongst them.
[If $a,b,c$ are all much smaller than
1, then $v_0\approx (b+c)/a$, so 
$\th_0\approx \coth^{-1}\bigl((b+c)/a\bigr)$.]
We infer that exotic behavior, including singularities
or runaway solutions, may typically occur
for aether boost angles of order unity.

\subsection{Restriction to physically viable couplings}

There are three independent coupling constants
that affect the solutions we are studying in either Einstein-aether 
theory or Horava gravity,
but stability and observational constraints 
restrict the range of physically viable values.
 
\subsubsection{Einstein-aether couplings}
As summarized in Ref.~\cite{Jacobson:2008aj}, 
$c_2$ and $c_4$ should be determined by 
$c_1$ and $c_3$ such that the preferred frame
parametrized post-Newtonian (PPN) 
parameters $\a_{1,2}$ vanish (or are very small compared to unity). 
Moreover, when $\a_{1,2}$ vanish, stability and positive energy
of linearized modes and the absence of vacuum Cherenkov
radiation by ultra high energy cosmic rays 
require $0<c_+<1$ and $0<c_-<c_+/(3(1-c_+))$,
where 
\beq
c_\pm=c_1\pm c_3.
\eeq
(In terms of $a,b,c$ these 
conditions correspond to $0<b<1$, $0<a<2b/(4-3b)$, and $c=-(a+b)/3$,
or, equivalently, $0<b<1$, $b(b-2)/(4-3b) <c<-b/3$, and $a=-b-3c$.)
We shall label the examples by their $c_\pm$ values.
In particular, the vacuum Cherenkov constraint requires
that all the mode speeds be greater than or equal to unity, so
except in the case where they are exactly unity, there
are values of the tilt angle where the dynamics is degenerate.
The degeneracy at $\th_2$ 
is relevant to the dynamics
only if $\dot{\sigma}_-\ne0$.

The remaining observational
constraint one might apply is that the radiation damping rate for
a binary pulsar system agrees
with the rate in GR, which agrees with observations within the
present relative uncertainty of about $0.002$.
The results of Ref.~\cite{Foster:2007gr} establish that 
this constraint is satisfied for generic small values
$c_i\lessim 0.001$, and if $c_i\lessim 0.01-0.1$ 
it is satisfied if $c_-\approx 0.18\, c_+$. 
For these values
the spin-0 mode speed is $v_0\approx 1.36$ and
the critical boost angle is $\theta_0\approx 0.94$.


\subsubsection{Horava gravity couplings}
The constraints on the couplings in Horava gravity
are the same as in Einstein-aether theory except for 
the PPN constraints $\a_{1,2}=0$,
which now are equivalent to $a=2b$~\cite{Blas:2010hb}.
The other constraints are $0<b<1$ and 
$(b+c)/(b(2+b+3c))>1$. Given the first of these, 
the second is satisfied in two
regions: (I) $c>(b+b^2)/(1-3b)$, $b<1/3$, and 
(II) $c<-2/3-b/3$, $c>(b+b^2)/(1-3b)$ when $b>1/3$.
The radiation damping rate has not yet been calculated
in the Horava case.

\subsection{Phase portraits}

As discussed in Sec. \ref{zeropm},
in the case when $p_\pm=0$  
one can reduce the dynamics to the
$\theta$ degree of freedom. Then
the dynamics can be displayed  
as a phase portrait in the 
$(\th,\dot\th)$ plane, 
exhibiting the flow of the vector
field $(\dot\th,\ddot\th)$.
This serves to illustrate the
general features of the dynamics 
discussed above.
\begin{figure}[thb!]
 \includegraphics[width=7.5cm]{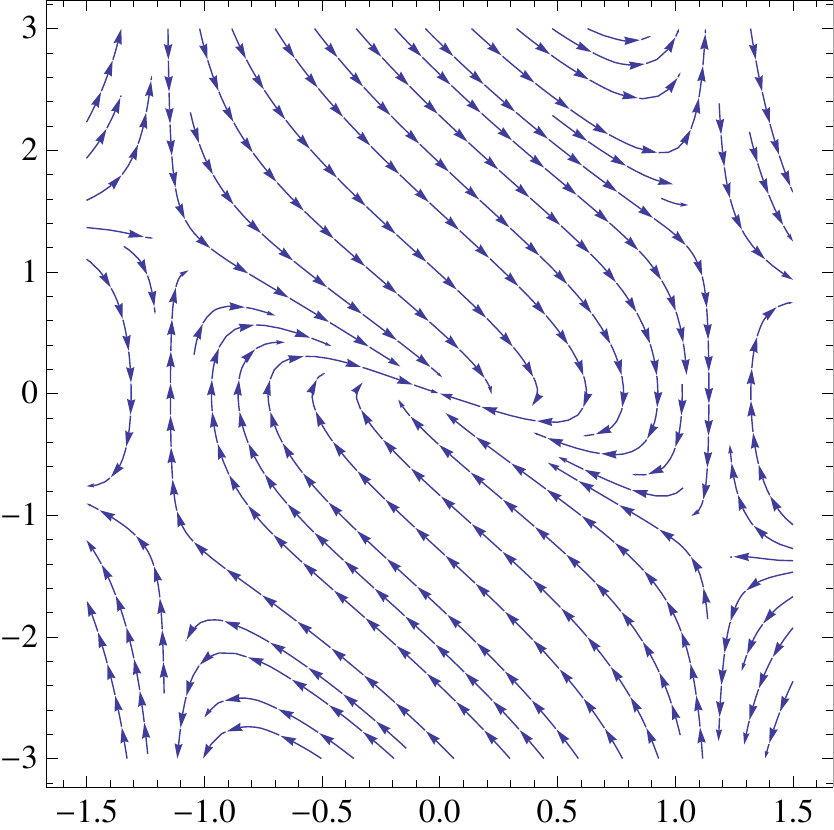} \\
\caption{\label{portrait110140}
Stream plot of the vector field $(\dot\th,\ddot\th)$
on the $(\th,\dot\th)$ plane, with $c_+=1/10$ and $c_-=1/40$,
and $p_\pm=0$.
In this case the determinant of $H_{ij}$ 
vanishes at $\th_0\simeq 1.16$.
}
\end{figure}
\begin{figure}[thb!]
 \includegraphics[width=7.5cm]{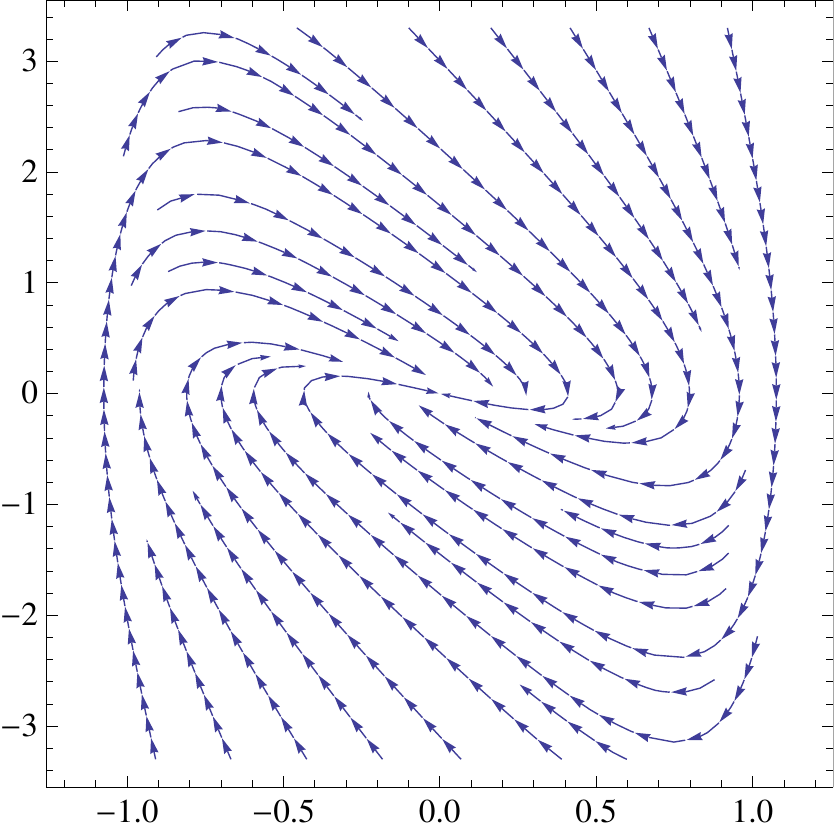} \\
\caption{\label{portrait1214}
Stream plot with $c_+=1/2$ and $c_-=1/4$,
and $p_\pm=0$.
In this case $\th_0\simeq 1.3$, but already 
for $\th\gtrsim 1.1$ 
and $\dot\th=0$ there is
no solution to the constraint equation for $\dot\a$.}
\end{figure}

Two examples are shown in Figs. 
\ref{portrait110140} and \ref{portrait1214}. 
In both of these the parameters $c_2$ and $c_4$ 
are chosen so as to satisfy the PPN constraints of
Einstein-aether theory,
and $c_\pm$ satisfy the remaining constraints
other than that of gravitational radiation damping. 
Figure \ref{portrait110140}  is qualitatively
similar to the phase 
portrait for the case $c_+=1/10$, $c_-=0.18 c_+$,
which is at least close to satisfying the radiation damping 
constraint. It is also similar to the case 
$(a,b,c)=(2/10,1/10,3/10)$ which satisfies the Horava constraints
other than the (unknown) radiation damping one.

In the case illustrated by 
Fig.~\ref{portrait110140},  runaway behavior 
occurs if $\th$ or $\dot\th$ is sufficiently large.
Numerical evolution of this example
suggests that
some of the flow lines end at curvature singularities.
The simplest spacetime scalar in this setting is the 
expansion of the aether, $\nabla_a u^a = 3\dot\a\cosh\th
+\dot\th\sinh\th$. For example, 
with initial data $\th=0$ and
$\dot\theta=4$, the evolution runs away to large
$\dot\th$. For another example,
with initial data $\th=1.5$ and
$\dot\theta\simeq1.8$, the evolution runs away to large
$\dot\a$. 

The singular behavior seen in these solutions
might be related to 
what is seen in some homogeneous anisotropic cosmologies with tilted 
perfect fluid matter (and vanishing cosmological constant)\cite{1979PhR....56...65C}.
It may appear inconsistent with the cosmological ``no-hair" theorem 
proved by Wald\cite{Wald:1983ky},
which showed that in the presence of a positive cosmological constant $\Lambda$,
all expanding Bianchi-type cosmologies (except Type IX) 
evolve toward the de Sitter solution with time scale
$\sqrt{3/\Lambda}$. But 
that result assumed that the dominant and strong energy 
conditions hold for the matter stress tensor. These conditions do not 
generally hold for the stress tensor associated with the aether
part of the action (\ref{Lae}).

\section{Conclusion}
The question driving this investigation was whether it  is natural 
for the aether to be aligned with the isotropic frame of a 
homogeneous, isotropic cosmology in Einstein-aether theory or Horava gravity? 
We addressed this question by studying the dynamics of a tilted
aether in a homogeneous anisotropic Bianchi type I cosmology 
with a cosmological constant. We found that generically the aether does align
provided its tilt angle and the time derivative of its tilt angle in units of the cosmological
constant are smaller than something of order unity. 
This extends the linearized stability result of KS~\cite{Kanno:2006ty}
to a finite basin of attraction whose precise shape
depends on the coupling parameters of the theory, 
and in some cases the basin appears to
be much broader than order unity. Outside of this basin, the solutions exhibit 
runaway or singular behavior of one or more of the variables.
Some of this behavior occurs when the propagation cone
of either the spin-0 or spin-2 mode is tilted enough to meet the
homogeneous constant $t$ surface. We do not know whether similar
behavior would persist if the homogeneous symmetry condition were dropped.

Our findings show that the fate of a universe with Bianchi I symmetry 
depends heavily on the initial tilt of the aether.
Perhaps the question of the 
initial tilt  could be addressed from the standpoint of 
quantum cosmology, for example via the ``wave function of the universe"
or via the distribution of initial conditions for chaotic inflation.

\begin{acknowledgments}
We thank John Barrow, William Donnelly, and 
Sugumi Kanno for helpful comments. 
This research
was supported in part by the NSF under Grant No. PHY-0903572. \\
\end{acknowledgments}

\bibliographystyle{utphys}
\bibliography{aether}

\providecommand{\href}[2]{#2}\begingroup\raggedright\begin{thebibliography}{10}

\bibitem{Jacobson:2008aj}
T.~Jacobson, ``{Einstein-aether gravity: a status report},'' {\em PoS} {\bf
  QG-PH} (2007)  020,
\href{http://arxiv.org/abs/0801.1547}{{\tt arXiv:0801.1547 [gr-qc]}}.

\bibitem{Horava:2009uw}
P.~Ho\v{r}ava, ``{Quantum Gravity at a Lifshitz Point},''
  \href{http://dx.doi.org/10.1103/PhysRevD.79.084008}{{\em Phys. Rev.} {\bf
  D79} (2009)  084008},
\href{http://arxiv.org/abs/0901.3775}{{\tt arXiv:0901.3775 [hep-th]}}.

\bibitem{Blas:2009qj}
D.~Blas, O.~Pujol\`{a}s, and S.~Sibiryakov, ``{Consistent Extension of
  Ho\v{r}ava Gravity},''
  \href{http://dx.doi.org/10.1103/PhysRevLett.104.181302}{{\em Phys. Rev.
  Lett.} {\bf 104} (2010)  181302},
\href{http://arxiv.org/abs/0909.3525}{{\tt arXiv:0909.3525 [hep-th]}}.

\bibitem{Lim:2004js}
E.~A. Lim, ``{Can we see Lorentz-violating vector fields in the CMB?},''
  \href{http://dx.doi.org/10.1103/PhysRevD.71.063504}{{\em Phys. Rev.} {\bf
  D71} (2005)  063504},
\href{http://arxiv.org/abs/astro-ph/0407437}{{\tt arXiv:astro-ph/0407437}}.

\bibitem{Li:2007vz}
B.~Li, D.~Fonseca~Mota, and J.~D. Barrow, ``{Detecting a Lorentz-Violating
  Field in Cosmology},''
  \href{http://dx.doi.org/10.1103/PhysRevD.77.024032}{{\em Phys. Rev.} {\bf
  D77} (2008)  024032},
\href{http://arxiv.org/abs/0709.4581}{{\tt arXiv:0709.4581 [astro-ph]}}.

\bibitem{ArmendarizPicon:2010rs}
C.~Armendariz-Picon, N.~F. Sierra, and J.~Garriga, ``{Primordial Perturbations
  in Einstein-Aether and BPSH Theories},''
  \href{http://dx.doi.org/10.1088/1475-7516/2010/07/010}{{\em JCAP} {\bf 1007}
  (2010)  010},
\href{http://arxiv.org/abs/1003.1283}{{\tt arXiv:1003.1283 [astro-ph.CO]}}.

\bibitem{Kanno:2006ty}
S.~Kanno and J.~Soda, ``{Lorentz violating inflation},''
  \href{http://dx.doi.org/10.1103/PhysRevD.74.063505}{{\em Phys. Rev.} {\bf
  D74} (2006)  063505},
\href{http://arxiv.org/abs/hep-th/0604192}{{\tt arXiv:hep-th/0604192}}.

\bibitem{Jacobson:2010mx}
T.~Jacobson, ``{Extended Ho\v{r}ava gravity and Einstein-aether theory},''
  \href{http://dx.doi.org/10.1103/PhysRevD.81.101502}{{\em Phys. Rev.} {\bf
  D81} (2010)  101502},
\href{http://arxiv.org/abs/1001.4823}{{\tt arXiv:1001.4823 [hep-th]}}.

\bibitem{Eling:2006df}
C.~Eling and T.~Jacobson, ``{Spherical Solutions in Einstein-Aether Theory:
  Static Aether and Stars},''
  \href{http://dx.doi.org/10.1088/0264-9381/23/18/008}{{\em Class. Quant.
  Grav.} {\bf 23} (2006)  5625--5642},
\href{http://arxiv.org/abs/gr-qc/0603058}{{\tt arXiv:gr-qc/0603058}}.

\bibitem{Blas:2010hb}
D.~Blas, O.~Pujolas, and S.~Sibiryakov, ``{Models of non-relativistic quantum
  gravity: the good, the bad and the healthy},''
\href{http://arxiv.org/abs/1007.3503}{{\tt arXiv:1007.3503 [hep-th]}}.

\bibitem{Mattingly2001}
D.~Mattingly and T.~Jacobson, ``Relativistic gravity with a dynamical preferred
  frame,'' in {\em Proceedings of the Second Meeting on CPT and Lorentz
  Symmetry}, V.~A. Kostelecky, ed.
\newblock (World Scientific, Singapore, 2002).
\newblock \href{http://arxiv.org/abs/gr-qc/0112012}{{\tt arXiv:gr-qc/0112012}}.

\bibitem{Carroll:2004ai}
S.~M. Carroll and E.~A. Lim, ``{Lorentz-violating vector fields slow the
  universe down},'' \href{http://dx.doi.org/10.1103/PhysRevD.70.123525}{{\em
  Phys. Rev.} {\bf D70} (2004)  123525},
\href{http://arxiv.org/abs/hep-th/0407149}{{\tt arXiv:hep-th/0407149}}.

\bibitem{Foster:2007gr}
B.~Z. Foster, ``{Strong field effects on binary systems in Einstein-aether
  theory},'' \href{http://dx.doi.org/10.1103/PhysRevD.76.084033}{{\em Phys.
  Rev.} {\bf D76} (2007)  084033},
\href{http://arxiv.org/abs/0706.0704}{{\tt arXiv:0706.0704 [gr-qc]}}.

\bibitem{1979PhR....56...65C}
C.~B. {Collins} and G.~F.~R. {Ellis}, ``{Singularities in Bianchi
  cosmologies.},'' \href{http://dx.doi.org/10.1016/0370-1573(79)90065-6}{{\em
  Phys. Rep.} {\bf 56} (1979)  65--105}.

\bibitem{Wald:1983ky}
R.~M. Wald, ``{Asymptotic behavior of homogeneous cosmological models in the
  presence of a positive cosmological constant},''
\href{http://dx.doi.org/10.1103/PhysRevD.28.2118}{{\em Phys. Rev.} {\bf D28}
  (1983)  2118--2120}.

\end{thebibliography}\endgroup

\end{document}